\documentclass{article}
\usepackage{spconf,amsmath,graphicx,cite}

\usepackage{subfig}
\usepackage{bm}
\usepackage{float}
\usepackage{amsfonts}
\usepackage{diagbox}
\usepackage{color}
\usepackage{url}
\usepackage{amssymb}
\usepackage{stackengine}
\usepackage{mathtools}
\usepackage{graphicx}
\usepackage{microtype}

\usepackage{algorithmic}

\usepackage{algorithm}
\title{Train your classifier first: Cascade Neural Networks Training \\from upper layers to lower layers}
\name{Shucong Zhang $^1$, Cong-Thanh Do $^2$, Rama Doddipatla $^2$, Erfan Loweimi $^1$, Peter Bell$^1$ and Steve Renals $^1$ \thanks{SZ was supported by a PhD Studentship funded by Toshiba Research Europe Limited.}}
\address{$^1$ Centre for Speech Technology Research, University of Edinburgh, Edinburgh, UK \\ $^2$  Toshiba Cambridge Research Laboratory, Cambridge, UK }
\begin{document}
%\ninept
%
\maketitle
\begin{abstract}
Although the lower layers of a deep neural network learn features which are transferable across datasets, these layers are not transferable within the same dataset. That is, in general, freezing the trained feature extractor (the lower layers) and retraining the classifier (the upper layers) on the same dataset leads to worse performance. In this paper, for the first time, we show that the frozen classifier is transferable within the same dataset. We develop a novel top-down training method which can be viewed as an algorithm for searching for high-quality classifiers. We tested this method on automatic speech recognition (ASR) tasks and language modelling tasks. The proposed method consistently improves recurrent neural network ASR models on Wall Street Journal, self-attention ASR models on Switchboard, and AWD-LSTM language models on WikiText-2.
\end{abstract}
\section{Introduction}
\label{Introduction}
The lower layers (close to the input) of a deep neural network (DNN) can be interpreted as a feature extractor while the upper layers (close to the output) can be viewed as a classifier. The feature extractor learns general low-level features which could be transferable across datasets  \cite{alexnet, yosinski2014transferable}. Transfer learning exploits this property -- in general, the feature extractor, which is trained on a base dataset, is transferred to a target dataset with the classifier retrained on the target dataset \cite{bengio2012transfer,swietojanski2012unsupervisedtf,ghoshal2013multilingualtf,cho2018multilingualtf}. Since the feature extractor learns general low-level features, the trained feature extractor should also be transferable within the same dataset. Therefore, retraining the classifier based on a frozen trained feature extractor using the same dataset should lead to little change of the model's performance. However, surprisingly, this retraining usually results in a performance drop \cite{yosinski2014transferable}.

% new paragraph
In this work, experimentally, we observe that if the classifier is trained with more data, when transferred to unseen datasets, it leads to better generalisation  (which is consistent with the transfer learning of feature extractors \cite{bengio2012transfer,swietojanski2012unsupervisedtf,ghoshal2013multilingualtf,cho2018multilingualtf}). Furthermore, we observed retraining the feature extractor based on the frozen trained classifier within the same dataset usually does not lead to performance drops but results in better generalisation .

%We investigate whether the upper layers of DNNs are transferable \textit{within the same dataset} -- can the feature extractor be retrained to fit a frozen trained classifier when retraining on the same dataset?   Experimentally, we observe that retraining based on frozen classifiers leads to better generalisation , indicating that the classifier is transferable on the same dataset. Therefore, although it is difficult for the classifier to learn good classification during retraining with a frozen feature extractor, the feature extractor can be retrained to provide suitable features which fit the the partition of the space defined by the frozen classifier.

Based on these observations, We propose a \textit{top-down}, layer-wise training method in which a network is trained by freezing the trained upper layers and retraining the lower layers in a cascade. We have applied this approach to train models for automatic speech recognition (ASR) using the clean Wall Street Journal (WSJ) dataset \cite{paul1992wsj}, the telephone conversation Switchboard (SWBD) \cite{godfrey1992switchboard} dataset and for language modelling (LM) using  WikiText-2 dataset. The trained models include long short-term memory (LSTM) \cite{hochreiter1997long} based connectionist temporal classification (CTC) \cite{graves2006ctc}, LSTM based  CTC-attention \cite{kim2017joint}, Transformer \cite{vaswani2017attention} and AWD-LSTM \cite{merity2018awdLM}. We observed performance gains over the conventional training where all the layers are trained jointly; to the best of our knowledge, this is the first time that a layer-wise training method consistently outperforms conventional joint training across different domains, models and tasks (ASR and LM).

In summary, we present three novel contributions:

\begin{itemize}
\vspace{-3mm}

\item Demonstration that the upper layers of neural networks are transferable within the same dataset;
\vspace{-3mm}

\item A top-down, layer-wise,  training algorithm; 
\vspace{-3mm}

\item Experiments indicating that top-down training  outperforms conventional joint training across different domains and models.

\end{itemize}

\vspace{-5mm}

\section{Related Work}
\vspace{-2mm}

Top-down training method can be viewed from different aspects. It is related to a variety of previous works, which we review in this section.

Layer-wise training without a joint fine-tuning stage has been investigated recently by a number of researchers  \cite{hettinger2017forward,  zhao2018retraining, mostafa2018deep, marquez2018deep, belilovsky2019greedy, nokland2019local}. However, unlike our top-down training algorithm, these training methods build DNNs in a bottom-up manner. In general, the reported experimental results do not indicate that these training methods surpass joint training. Belilovsky et al~\cite{belilovsky2019greedy} and Nokland et al~\cite{nokland2019local} have reported layer-wise training outperforms joint training for training convolutional neural networks for image classification. However, these methods are not effective for training networks with residual connections. 

Zhang et al~\cite{zhang2020learning} propose to use the classifier trained on the clean data to train the feature extractor on noisy data, so the noisy feature extractor is forced to learn features which fit the clean classifier and thus the method has the effect of denoising. In this work we show the classifier is transferable not only across different datasets but also within the same dataset. 

Both the proposed training method and dropout \cite{srivastava2014dropout} reduce the size of the network during training. Dropout can be viewed as reducing the effective width of the network in a random manner, while our training method reduces the effective depth of the network by following a fixed schedule. When these two methods are combined, during training thinner and shallower sub-networks are trained while during testing, the model has the representation power of a wide and deep network. In our experiments, the combination of these two methods yields significant performance gains. 

In  gradient based optimization, the learning rate can either be dynamically tuned by the user or be tuned by optimization algorithms \cite{qian1999momentum,duchi2011adaptive,zeiler2012adadelta,kingma2014adam,dozat2016incorporating}. In top-down training, since the classifier is frozen, its learning rate  can be viewed as 0 when retraining. Our experiments show that this procedure is compatible with popular methods of changing the learning rate dynamically.

\vspace{-2 mm}
% new section
\section{Transferable Upper Layers}
\label{sec: transferable upper layers}
\vspace{-2 mm}

In this section, we empirically show: the quality of the frozen classifier decides the quality of the retrained model and in the joint training the quality of the classifier increases then decreases. We present the top-down training algorithm in Section~\ref{sec:algorithm} and present comprehensive comparisons between the top-down training method and joint training in Section~\ref{experiments}.

We define \textit{general classifier} by making an analogy to the concept of general feature extractor: if the classifier is more general, then training the feature extractor with respect to it on unseen data should lead to better generalisation compared to a classifier which is not general. We hypothesize that with more training data, the trained classifier becomes more general/has higher quality. To verify this, we conducted a series of end-to-end speech recognition experiments using the WSJ. We divide the WSJ training set into 5\%, 10\%, 20\%, 40\% and 80\% size subsets, leaving 20\% unseen training data. We train 4-layer BLSTM CTC models and the training stops if for 5 consecutive epochs there is no improvement on the development set (dev93). Other experimental setups follow \cite{kim2017joint}. We group the softmax layer and the topmost BLSTM layer as the classifier.

\begin{figure}[tb!]
\centering
{\includegraphics[width=1.0\columnwidth]{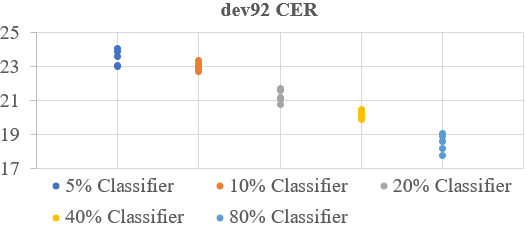}}
\quad \newline
{\includegraphics[width=1.0\columnwidth]{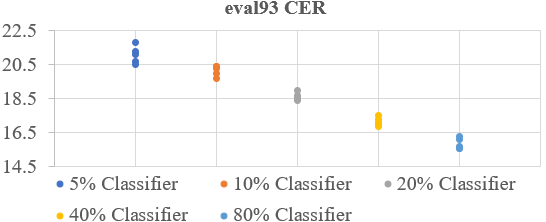}}

\caption{The CERs of models trained on the 20\% unseen data with frozen classifiers from trained models on different si284 subsets. The experiments are repeated 5 times with different random seeds. } 
\label{fig:subclassifier}
\end{figure}

\begin{table}[h]
\centering
\caption{CERs of models trained on si284 subsets.}
\vspace{-2 mm}
\label{tab:subclassifier}
\begin{tabular}{l|l}
 \hline
Training set                                     & dev92/eval93 (CER) \\ \hline
5\% train si284                       & 34.4/32.9      \\
10\% train si284                      & 28.3/26.1       \\
20\% train si284                      & 22.5/19.9       \\
40\% train si284                      & 17.5/14.7       \\
80\% train si284                     & 13.2/10.6       \\ \hline
\end{tabular}
\end{table}

As shown in Table~\ref{tab:subclassifier}, with more training data, the model achieves better ASR performance. To test if the classifiers from each trained model become more general as the size of the training subsets increases, on the unseen 20\% training set, we run experiments of training the feature extractor with these frozen trained classifiers. Figure~\ref{fig:subclassifier} shows on the unseen training data, training with respect to the frozen classifier trained on larger subsets consistently yields lower character error rates (CERs). Thus, we conclude the classifiers trained on larger subsets are indeed more general/have higher quality, and training feature extractor with them leads to better generalisation.

%We also train a CTC system on the 20\% subset. Table~\ref{tab:20} shows the results. The model trained on the 20\% subset yields inferior results compared to the model trained on the 20\% unseen dataset. However, when we take the classifier from the inferior model and train the feature extractor with respect to it, it consistently yields better CERs compared to the better model which is conventionally trained on the 20\% unseen dataset. 

%\begin{table}[h]
%\centering
%\caption{CERs of models trained on different 20\% si284 subsets. ``Joint train'' means all the layers are trained jointly and ``From A'' indicates the classifier is from the model trained on A and it is frozen. For experiments with multiple runs, the CER is the average with one std. }
%%\label{tab:20}
%\begin{tabular}{l|l|l}
% \hline
%Training set     & Classifier      & dev93/eval92 (CER) \\ \hline
%20\% si284 (A)      & Joint Train               & 22.5/19.9       \\
%20\% Unseen (B)     & Joint Train             & 22.1/19.3       \\
%20\% Unseen (B)    & From A              & 21.3$\pm$0.3 /18.6$\pm$0.2       \\ %\hline
%\end{tabular}
%\end{table}

\begin{figure}[h]
\centering
\subfloat[Loss on 80\% joint training]{\includegraphics[width=0.40\columnwidth]{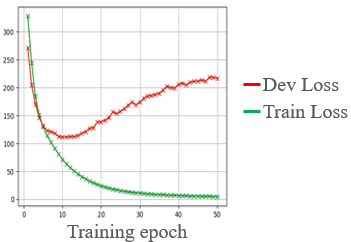}}
\quad
\subfloat[CERs on 20\% unseen ]{\includegraphics[width=0.45\columnwidth]{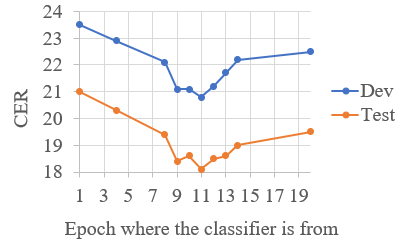}}
\qquad

\caption{Train/dev loss for the joint training  on the 80\% subset and CERs of the models trained on the 20\% unseen data with frozen classifiers taken from the epochs of the joint training. } 
\label{fig:training loss}
\end{figure}

We further investigate in the joint training, does the classifier become more general as the training proceeds? We train the CTC model on the 80\% subset for a sufficiently large number of epochs. Then, we take the classifiers from the models among these training epochs and retrain the feature extractor based on them respectively on the 20\% unseen data. Figure~\ref{fig:training loss} shows on the 20\% unseen data, the model trained with the frozen classifier which is from epoch 11 of the joint training on the 80\% subset gives the lowest CERs. In the joint training on the 80\% subset, the model around epoch 11 also gives the lowest development loss. On the 20\% unseen data, the models trained with frozen classifiers which are from later epochs of the joint training on the 80\% subset have increasing CERs. Also, in the joint training on the 80\% subset, the development loss increases after epoch 11. Thus, the classifier firstly becomes general (low CERs when transferred to the unseen data), then it becomes overfitting and leads to worse generalisation  (high CERs when transferred to the unseen data). 

If we freeze the classifier when it is general/of high-quality, we can prevent it from becoming overfitting and we can further train the feature extractor with the high-quality classifier. Based on this, we propose the top-down training algorithm, which is described in Section~\ref{sec:algorithm}. The opposite strategy (freezing the feature extractor then retraining the classifier on the same dataset) usually leads to higher error rates \cite{yosinski2014transferable}.

\vspace{-3 mm}

\section{Top-down Training Algorithm}
\label{sec:algorithm}

\vspace{-2 mm}

Based on the findings of Section~\ref{sec: transferable upper layers}, we develop a novel top-down, layer-wise training algorithm, which can be viewed as a process of searching for high-quality classifiers.  The algorithm searches among  models trained by  conventional joint training along the epoch dimension and the layer dimension (layer index).

\begin{algorithm}[h]
   \caption{Greedy Layer-wise Top-down Training }
   \label{alg:top-down}
\begin{algorithmic}
   \STATE {\bfseries Input:} A trained $n$ layer network $M$
   \STATE $e \gets$ validation error of $M$ 
   \FOR{ $i=1$ {\bfseries to} $n-1$ }
   \STATE $M' \gets M$
   \STATE Freeze top $i$ layers (near output) of $M'$
   \STATE Reinitialize bottom $n-i$ layers of $M'$
   \STATE Retrain bottom $n-i$ layers of $M'$
   \STATE $e' \gets$ validation error of $M'$ 
   \IF{$e' > e$}
   \STATE BREAK
   \ENDIF
   \STATE $M \gets M'$
   \STATE $e \gets e'$ 
   \ENDFOR
   \STATE {\bfseries Output:} A retrained n-layer network $M$
\end{algorithmic}
\end{algorithm}

Searching the time/epoch dimension is useful since at the beginning of the training the classifier is  under-trained and at the end of the training the classifier tends to be overfitting. 
%Suppose it takes $N$ epochs for the training to converge. 
We denote the model at epoch $p$ as $M^p$. We search for the best classifier through all $M^p$.

Searching the layer dimension is essential since arbitrarily many upper layers can be grouped and viewed as the classifier (the feature extractor should at least have one layer). For a $n$-layer neural network $M$ which is trained by the joint training, freezing the top $i$ layers of it then reinitializing and retraining the bottom $n-i$ layers generates a new model $M_{i,n-i}$. For the model $M_{i,n-i}$, we can add newly retrained lower layers to the classifier and retrain the remaining lower layers. $M_{i,j,n-i-j}$ denotes a model trained by freezing the top $i+j$ layers of model $M_{i,n-i}$ followed by retraining the lower $n-i-j$ layers. This corresponds to a search of  ordered sequences of natural numbers $(i,j,k,\cdots,r)$ such that $i+j+k+\cdots+r=n$; for a network $M$ with three layers, the search would be across  $M_{1,1,1}$, $M_{1,2}$ and $M_{2,1}$ (e.g., $M_{1,1,1}$ indicates train all the layers jointly;then freeze the topmost layer  and retrain the bottom two layers; then freeze the top two layers and retrain the bottom most layer).

As the complete search along these two dimensions is expensive, we use a greedy search algorithm for layer-wise training (Algorithm~\ref{alg:top-down}),  which uses a converged model for the epoch dimension. For the layer dimension, we freeze layers from top to bottom in a layer-by-layer manner (i.e. we only consider $M_{1,n-1}, M_{1,1,n-2}, \cdots M_{1,1,\cdots,1}$). We also use the validation set to halt the search. The complexity of this algorithm is $O(n)$, where $n$ denotes the number of layers of the network. We omit the complexity of training the network since it is independent of the top-down training algorithm. 

%We use this greedy search in most of our experiments. We  execute part of the complete search in some experiments to show if it is beneficial, especially when  greedy search fails (i.e., in the first round of retraining the classifier should have more than one layer or the classifier should be taken from an early epoch). 

\vspace{-2 mm}
\section{Experiments}
\label{experiments}
\vspace{-2 mm}

We apply the top-down training algorithm 
%(part of the complete search and the greedy search)
to train a range of neural networks on speech and text domains. %The models and datasets used are  summarized in Table~\ref{tab:experiment summary}. 
We have used end-to-end speech recognition models. We test the end-to-end models rather than the conventional hybrid system since the end-to-end models have only one feature to output unit component, allowing an exact fair comparison between the joint and top-down training. For a hybrid system,  better training of one component may not lead to a better performance of the entire system. For the same reason, we do not perform language model fusion or adaptation. All models were built using PyTorch \cite{paszke2017pytorch}. 

\iffalse
\begin{table}[h]
\caption{Tasks and models used for top-down training.}
\vspace{-2 mm}
\centering
\label{tab:experiment summary}
\begin{tabular}{l|l|l}
\hline
Task                                                                 & Dataset    & Model                                                        \\ \hline
Speech Recognition & WSJ        & \begin{tabular}[c]{@{}l@{}}CTC \\ CTC-Attention\end{tabular} \\ \hline
Speech Recognition                                                     & SWBD   & \begin{tabular}[c]{@{}l@{}}Transformer\end{tabular}  \\ \hline
Language modelling setup                                                      & WikiText-2 & AWD-LSTM                                                     \\ \hline
\end{tabular}
\end{table}
\fi

\vspace{ -3 mm}

\subsection{Experiments on Speech Recognition}
\vspace{ -1 mm}
We employ our method to train CTC models on the full WSJ si284 training set. The experimental setup follows \cite{kim2017joint}. We used three different random seeds to build three baselines, and executed  greedy  top-down layer-wise training (Algorithm~\ref{alg:top-down}) for each baseline. For seed 1, we also tested freezing the lowest layer, followed by reinitializing and retraining the remaining upper layers. 

\begin{table}[!h]
\caption{CERs for experiments on WSJ.}
\vspace{-2mm}
\label{tab:wsj}
\centering
\begin{tabular}{l|l|l|l}
\hline
 model     & \multicolumn{3}{l}{dev93/eval92 (CER)}              \\ \hline 
CTC     & seed1            & seed 2           & seed 3           \\ \hline 
Baseline     & 12.4/9.7          & 12.6/10.4          &12.6/10.2          \\
+ top-down training  & 10.8/8.2          & 10.6/8.5          &     11.5/8.9     \\
+ freeze lowest  & 13.1/10.5 & \_          & \_ \\
\hline
\multicolumn{3}{l|}{ Dropout 0.5} & \multicolumn{1}{l}{9.6/7.6}                  \\ 
\multicolumn{3}{l|}{ + top-down training } & \multicolumn{1}{l}{\textbf{8.2/6.3}}                  \\ \hline
\multicolumn{3}{l|}{ CTC-attention} & \multicolumn{1}{l}{seed 1}                  \\  \hline
\multicolumn{3}{l|}{ Baseline A} & \multicolumn{1}{l}{8.0/6.2}                  \\ 
\multicolumn{3}{l|}{ + freeze softmax layers from A} & \multicolumn{1}{l}{\textbf{7.3/5.4}} \\
\multicolumn{3}{l|}{ + freeze softmax layers from B} & \multicolumn{1}{l}{7.5/5.7} \\ \hline
\multicolumn{3}{l|}{VGG CTC-attention} & \multicolumn{1}{l}{seed 1}                  \\  \hline
\multicolumn{3}{l|}{ Baseline B} & \multicolumn{1}{l}{7.2/5.4}                  \\ 
\multicolumn{3}{l|}{ + freeze softmax layers from B} & \multicolumn{1}{l}{7.0/5.2}                  \\ 
\multicolumn{3}{l|}{ + freeze softmax layers from A} & \multicolumn{1}{l}{\textbf{6.7/5.2}}                  \\ \hline \hline

\multicolumn{4}{l}{Previous works} \\ \hline  \hline
\multicolumn{3}{l|}{ CTC DNN \cite{hannun2014first}} & \multicolumn{1}{l}{\_ /10.0}                  \\ \hline
\multicolumn{3}{l|}{ CTC BLSTM \cite{graves2014towards}} & \multicolumn{1}{l}{\_ /9.2}                  \\ \hline
\multicolumn{3}{l|}{CTC-Attention  \cite{kim2017joint}} & \multicolumn{1}{l}{11.3/7.4}                  \\ 
\multicolumn{3}{l|}{+ pad silence \cite{zhang2019windowed}} & \multicolumn{1}{l}{7.8/5.8}                  \\ 
\multicolumn{3}{l|}{+ deep encoder \cite{delcroix2018auxiliary}} & \multicolumn{1}{l}{7.4/5.5}                  \\ 
\hline

\end{tabular}
\end{table}

Top-down training significantly reduces the CER for each of the baseline models (Table~\ref{tab:wsj}). For seed 1 and seed 2, the layer-wise training stops at the lowest layer. For seed 3 (where the proposed method brings the least improvement), the training procedure stops at the layer above the lowest layer (i.e. the final model is $M_{1,1,1,2}$). An additional experiment on seed 1, indicates that if we freeze the lowest layer and retrain the top layers, the accuracy of the model drops. This observation is also consistent with the findings of \cite{yosinski2014transferable}. 

Using seed 3, we train a model with dropout probability $0.5$ and then apply top-down training. We observe large gains in test accuracy  over both the baseline and the model trained with dropout ($38\%$ and $17\%$ relative gain, respectively). The combination of these two training methods performs impressively, with the CTC model giving CERs comparable to the joint CTC-attention baseline model, which is more flexible and has more capacity. 

We apply the proposed training method to train hybrid CTC-attention models \cite{kim2017joint} on WSJ. The experimental setup of BLSTM CTC-attention follows \cite{kim2017joint} and the setup of VGG BLSTM CTC-attention follows \cite{delcroix2018auxiliary}. We  freeze only the softmax layer of the CTC part and the softmax layer of the attention based encoder-decoder part (the final model is $M_{1,n-1}$) due to the large number of experiments; and performance gains are already observed on $M_{1,n-1}$. Table~\ref{tab:wsj} demonstrates that, to the best of our knowledge, top-down training results in state-of-the art character error rates for LSTM-based end-to-end models on WSJ, without language model fusion or adaptations.

%The architectures of the decoder/attention parts in model A and model B are the same. However, model B has a  deeper encoder. Table~\ref{tab:wsj} also indicates that by executing a better training algorithm, a shallow model (model A) achieves similar performance to a deep model (model B).Besides retraining each model with its own trained frozen softmax layers, we also retrained model A based on the frozen softmax layers of model B and  vice versa. As shown in Table~\ref{tab:wsj}, retraining according to the softmax layers of model B leads to worse results than  retraining based on the softmax layers of model A. This indicates that, compared to the shallow baseline, the better performance of the deep baseline comes from the deep feature extractor (encoder), rather than the classifier. 

We test the proposed training method on SWBD. We employ the Transformer \cite{vaswani2017attention} based CTC-attention model and the experimental setup follows \cite{karita2019comparative}. We  freeze only the softmax layer of the CTC part and the softmax layer of the Transformer due to large number of experiments. Table~\ref{tab:swbd} shows the proposed method significantly reduces the word error rates (WERs) over the baseline, and gives noticeably better results compared to the previous works. 
\vspace{-2mm}
\begin{table}[t]
\caption{WERs for experiments on SWBD}
\vspace{-2mm}
\label{tab:swbd}
\centering
\begin{tabular}{l|l}
\hline
model & SWBD/Callhm (WER) \\ \hline
Baseline                   &  9.0/18.1                  \\ \hline
+ freeze softmax layers    &   \textbf{8.6/17.2}                \\ \hline \hline 
\multicolumn{2}{l}{Previous works} \\ \hline \hline 
Transformer \cite{karita2019comparative}                    &  9.0/18.1                 \\ \hline
Very deep self-attention \cite{pham2019veryDeep}                    &  10.4/18.6                 \\ \hline
Multi-stride self-attention \cite{han2019multi} &  9.1/\_  \\\hline
\end{tabular}
\end{table}

\vspace{-2 mm}
\subsection{Experiments on Language Modelling}
\vspace{-1 mm}

We apply the proposed method in training the AWD-LSTM language model \cite{merity2018awdLM} on WikiText-2 dataset. The architecture of the networks and the training procedure are same as in the previous work \cite{merity2018awdLM}. We run the retraining three times -- with the same/different initial weights as the baseline. We freeze the topmost layer and retrain the bottom layers. The input/output embedding are tied in both of the joint training and the retraining (so the input embedding is also frozen during the retraining). This does not contradict the top-down approach since although tied with the input embedding, the output embedding is still the closest to the output and in general the gains of the tied embedding is from the output embedding \cite{press2017tiedembedding}. The top-down method can be viewed as making the LSTM sufficiently trained by retraining the LSTM based on the trained and frozen word embedding. 

As shown in Table~\ref{tab:lm_wikitext}, although the performance gains from the top-down training are not as large as the gains in previous experiments, the reduction of the perplexity is consistent. The proposed training method may implicitly overlap with some regularization techniques used in the training of AWD-LSTM. However, these  regularization methods do not nullify the advantages of the top-down training.

\vspace{-2 mm}

\begin{table}[h!]
\caption{Perplexity of the AWD-LSTM language models on WikiText-2.}
\vspace{-2 mm}
\centering
\label{tab:lm_wikitext}
\begin{tabular}{l|ll}
\hline
Model                         & Dev           & Test          \\ \hline
Baseline seed 1             & 68.7          & 65.6          \\ \hline
+ freeze topmost layer seed 1 & 68.2          & 65.3          \\
+ freeze topmost layer seed 2 & 68.2          & \textbf{65.2} \\
+ freeze topmost layer seed 3 & \textbf{68.1} & \textbf{65.2} \\ \hline \hline
Previous work \cite{merity2018awdLM}       & 68.6          & 65.8          \\ \hline
\end{tabular}
\end{table}
 
\vspace{-3mm}
\section{Conclusion}
\vspace{-2mm}

In this paper, we have shown that upper layers are in general transferable. Based on this, we develop a novel layer-wise top-down training method, which prevents the upper layers from overfitting. We demonstrate, for the first time,  that layer-wise training can outperform conventional joint training across speech recognition and language modelling tasks. Future work will include reducing the complexity of the search. 
% References should be produced using the bibtex program from suitable
% BiBTeX files (here: strings, refs, manuals). The IEEEbib.bst bibliography
% style file from IEEE produces unsorted bibliography list.
% -------------------------------------------------------------------------
\newpage
\bibliographystyle{IEEEtran}

\bibliography{strings}

\end{document}